\documentclass[aps,prl,preprint,superscriptaddress]{revtex4-1}
\usepackage{graphicx}
\PassOptionsToPackage{unicode}{hyperref}
\PassOptionsToPackage{naturalnames}{hyperref}
\usepackage{natbib}
\usepackage{amssymb}
\usepackage{amsmath}
\usepackage{hyperref}
\usepackage{scalerel}
\usepackage{float}
\usepackage{mathtools}
\usepackage{dcolumn}
\begin{document}

\title{Anisotropy of the proton kinetic energy as a tool for capturing structural transition in nanoconfined H$_2$O\\}

\author{Mohd Moid}
\affiliation{Centre for Condensed Matter Theory, Department of Physics, Indian Institute of Science, Bangalore 560012, India}
\author{Yacov Finkelstein}
\affiliation{Nuclear Research Center - Negev, Beer-Sheva 84190, Israel}
\author{Raymond Moreh}
\affiliation{Physics Department, Ben-Gurion University of the Negev, Beer-Sheva 84105, Israel}
\author{Prabal K Maiti}
\email[Correspondence email address: ]{maiti@iisc.ac.in}
\affiliation{Centre for Condensed Matter Theory, Department of Physics, Indian Institute of Science, Bangalore 560012, India}

\begin{abstract}
The proton dynamics of a 2D water monolayer confined inside a graphene slit pore is studied in Cartesian and molecular frames of reference using molecular dynamics simulations. The vibrational density of states of the proton was calculated versus temperature and was further used to deduce the mean kinetic energy of the hydrogen atoms, Ke(H), in both frames of reference. The directional components of Ke(H) are in good agreement with experimental observations for bulk as well as nanoconfined water. Nonetheless, while in the molecular frame of reference the effect of temperature on the anisotropy ratios of Ke(H) (the ratio between its directional components) are practically invariant between the 2D and 3D cases, those in the Cartesian frame of reference reveal a rather notable reduction across 200K, indicating the occurrence of an order-disorder transition. This result is further supported by the calculated entropy and enthalpy of the confined water molecules. Overall it is shown that Ke(H) anisotropy ratios may serve as a valuable order parameter for detecting structural transformations in hydrogen bonds containing molecular systems.\par
\end{abstract}
\maketitle
Water in confined geometries exhibit intriguing behaviors, much different from the ordinary bulk, hence studies of nanoconfined water provide valuable information regarding the physico-chemical nature of water dynamics, e.g. freezing\cite{David2019}, fluid organization\cite{Mochizuki2015,Koga2001,Algara2015}, mobility\cite{Holt2006,Tunuguntla2017,Secchi2016,Mukherjee2007,Bonthuis2011,Naskar2020} and structural and dynamical transitions\cite{Noon2002,Bai2006,Nomura2017,Majumdar2021,Joseph2008}. Structural transitions of confined water strongly depend on the nature and scale of confinement\cite{Shiomi2007,Chakraborty2017,Mukherjee2008}. An interesting feature of confinement that receives a detailed attention in recent years, regards the fluid-solid interface, with the goal of understanding how the hydrophobicity and hydrophilicity of the confining lattice surface affects the structure and dynamics of guest water molecules\cite{Schlaich2016,kayal2015, Lupi2014}. An exemplifying and well-studied example is water inside single wall carbon nanotubes (SWCNTs)\cite{Mochizuki2015,Whitby2008, Falk2010}.  \par
Deep inelastic neutron scattering (DINS) is an exclusive experimental technique which is extensively used to explore proton dynamics in hydrogen bonds (HBs) containing systems. It allows for probing the dynamics of protons, as well as possible related nuclear (NQEs) and competing (CQEs) quantum effects, via a measurement of the proton momentum distribution and hence of its mean kinetic energy, Ke(H). This parameter was proved to be a fairly sensitive probe of H-dynamics, which by itself is dominated by the way the HBs network into three or reduced (two and one) dimensional phases. Up until now, the accumulated DINS findings suggest that when water is confined in 3D spaces of effective pore diameter less than 20{\AA}, Ke(H) assumes anomalous values compared to those occurring in ice and liquid bulks\cite{Reiter2012,Kolesnikov2016,Senesi2007,Pantalei2011}. In one of the DINS studies an attempt was made to probe the effect of 2D confinement on the H-dynamics in liquid (293K) and ice (20K), both confined between 12{\AA} separated hydrophilic surfaces of graphene oxide sponges (GOs) at ambient pressure\cite{Romanelli2015,Andreani2017}. However, the measured Ke(H) values [summarized in Table \ref{table}] did not reveal any evidence of Ke(H) anomaly as compared to regular bulk water phases\cite{Colognesi2011,Finkelstein2014,Flammini2012,Andreani2013}. Note that for bulk water, the value of Ke(H) remains practically constant between 5K and 300K due to the dominance of the zero point value of Ke(H) in water\cite{Finkelstein2014}. \par
\begin{figure}[!ht]
\begin{center}
\includegraphics[width=3.375in,keepaspectratio=true]{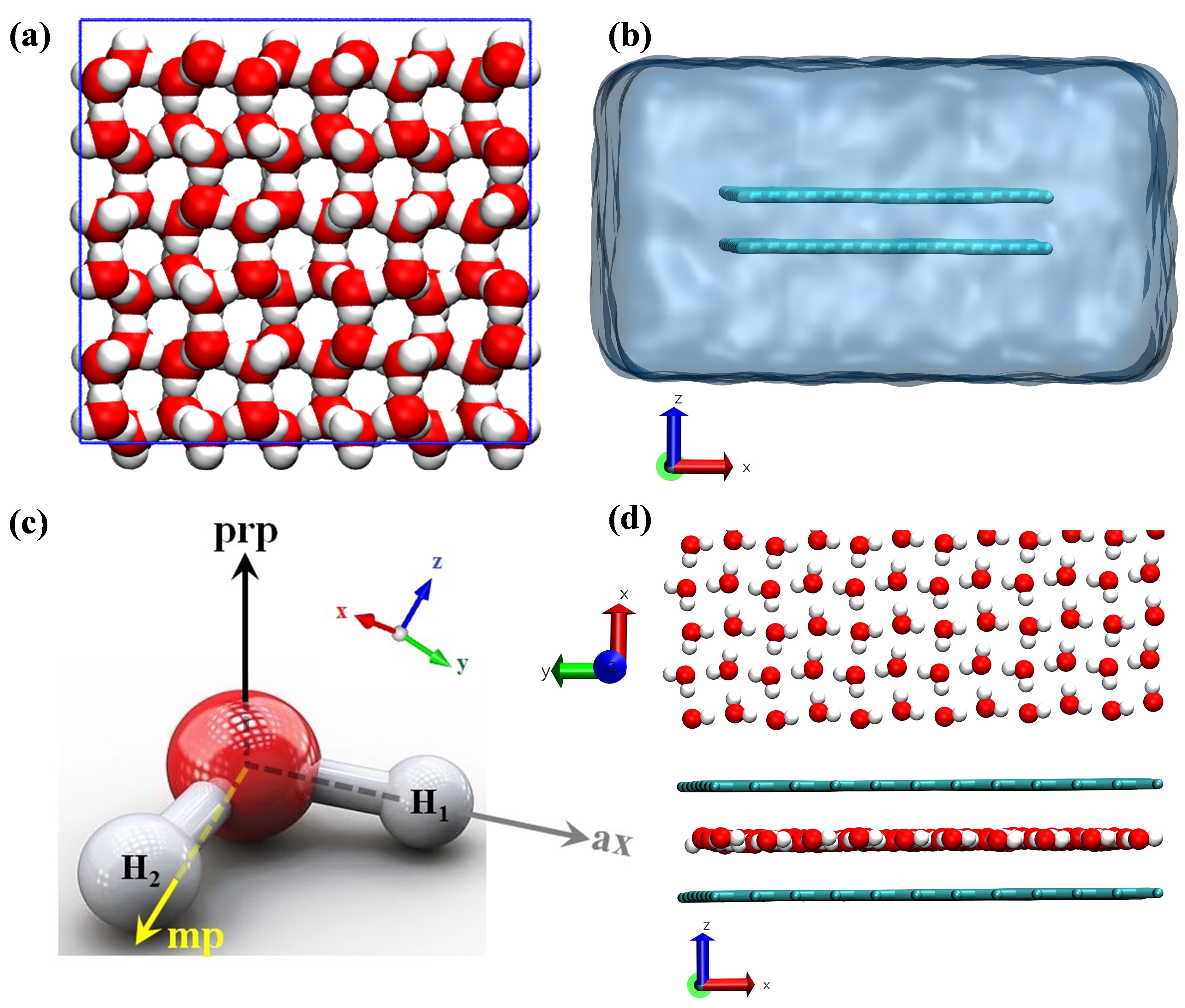}
\caption{The initial structure of a 3D hexagonal bulk ice (a) and of 2D square ice  confined inside a graphene slit-pore (b). Red and white colors denote oxygen and hydrogen atoms respectively. Gray lines and cyan background in (b) denote the graphene sheet and the water bath respectively. (c) the H$_2$O molecular frame of reference, as defined  for an arbitrary orientation of a water molecule relative to the Cartesian frame of reference: ax (grey) is directed along one of the OH bonds, mp (yellow) lays in the H$_2$O molecular plane but extends normal to ax, and prp (black) is perpendicular to the H$_2$O plane. The relative orientations of the sandwiched ice monolayer and the confining graphene layers is given in (d). Note that in the simulation the Cartesian directions are fixed whereas those of the molecular frame of reference are updated with time according to the reorientation of  molecules during the simulation.}
\label{system}
\end{center}
\end{figure}

The neutron scattered intensities in DINS experiments were also used to resolve the three Cartesian components of Ke(H), i.e. those along and perpendicular to the OH bond direction in 2D confined H$_2$O molecules\cite{Romanelli2015}. It was found that the directional components of Ke(H) show a similar anisotropy as those previously reported for bulk ice\cite{Flammini2012} and water\cite{Andreani2013} in the molecular frame of reference. The existing data is summarized in Table \ref{table} with the orthogonal directions in the molecular frame of reference defined in Fig. \ref{system}(c).\par
In recent years, the partial vibrational density of states (pVDOS) approach was utilized for studying proton dynamics in various H containing systems\cite{Flammini2012,Senesi2013,Pietropaolo2009,Andreani2016,Moid2020}, and is applied here for studying proton dynamics in nanoconfined water under 2D topology.\par  
Using this approach, we computed Ke(H) in two coordinate system, namely the Cartesian and Molecular frames of reference. To simulate the directional components of Ke(H) in any of the above coordinate system, we start by constructing the kinetic energy tensor whose principal components determine the kinetic energy values along the orthogonal direction of that coordinate system[see Appendix 1 for details].
For the simulation of both bulk and nanoconfined water phases, the flexible TIP4P-2005 water model\cite{Abascal2005,Gonzalez2011} was utilized. For the 2D ice structure\cite{Algara2015,Mario2015,Pascal2019}, simulations were performed using a 4x4 nm$^2$ graphene nanoslit pore of a 6.8 {\AA} separation distance, embedded in a water bath [see Appendix 2 for further details of the simulation].\par
\begin{figure}[!ht]
\begin{center}
\includegraphics[width=3.375in,keepaspectratio=true]{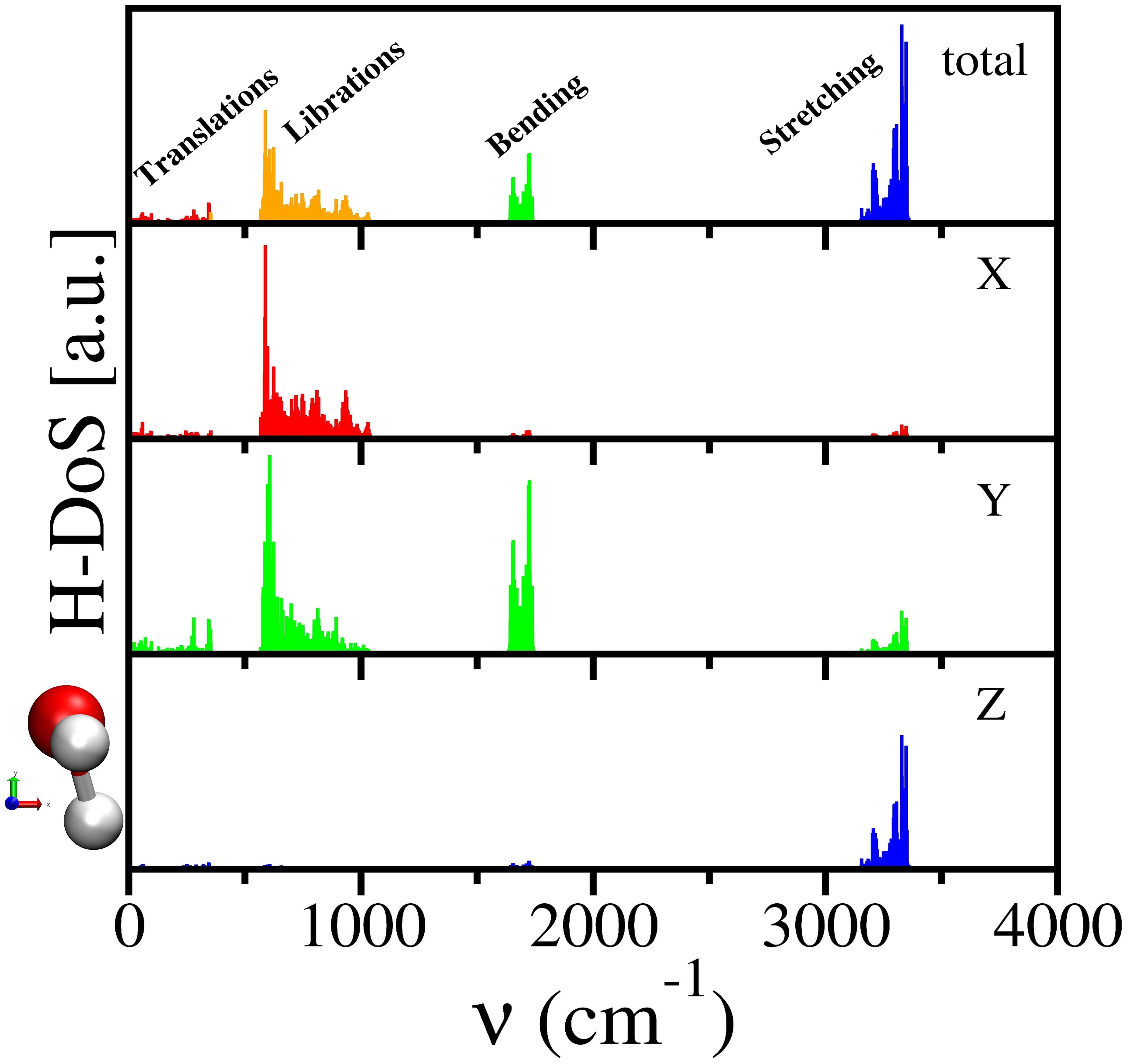}
\caption{The upper panel depict the total H-VDOS resulting from a 500ps trajectory, computed for bulk ice at 5 K.  The frequency bands at $\sim$[0-350] cm$^{-1}$ (red) and $\sim$[350-1250] cm$^{-1}$ (orange) comprise the lattice modes of translation and libration repectively. The peaks centered around $\sim$1700 (green) and 3300 (blue) cm$^{-1}$ arise from the molecular HOH bending and two OH stretching modes. The lower panels correspond to the computed projections in the molecular frame of reference of the H1-VDOS of an arbitrary H$_2$O molecule in bulk ice at 5K. Its individual orientation is depicted on the left. The noticeable differences between the three directional components of the H-VDOS illustrate the strong anisotropy of Ke(H) occurring in the molecular frame of reference.}
\label{DoS}
\end{center}
\end{figure}
\begin{figure}[!ht]
\begin{center}
\includegraphics[width=3.375in,keepaspectratio=true]{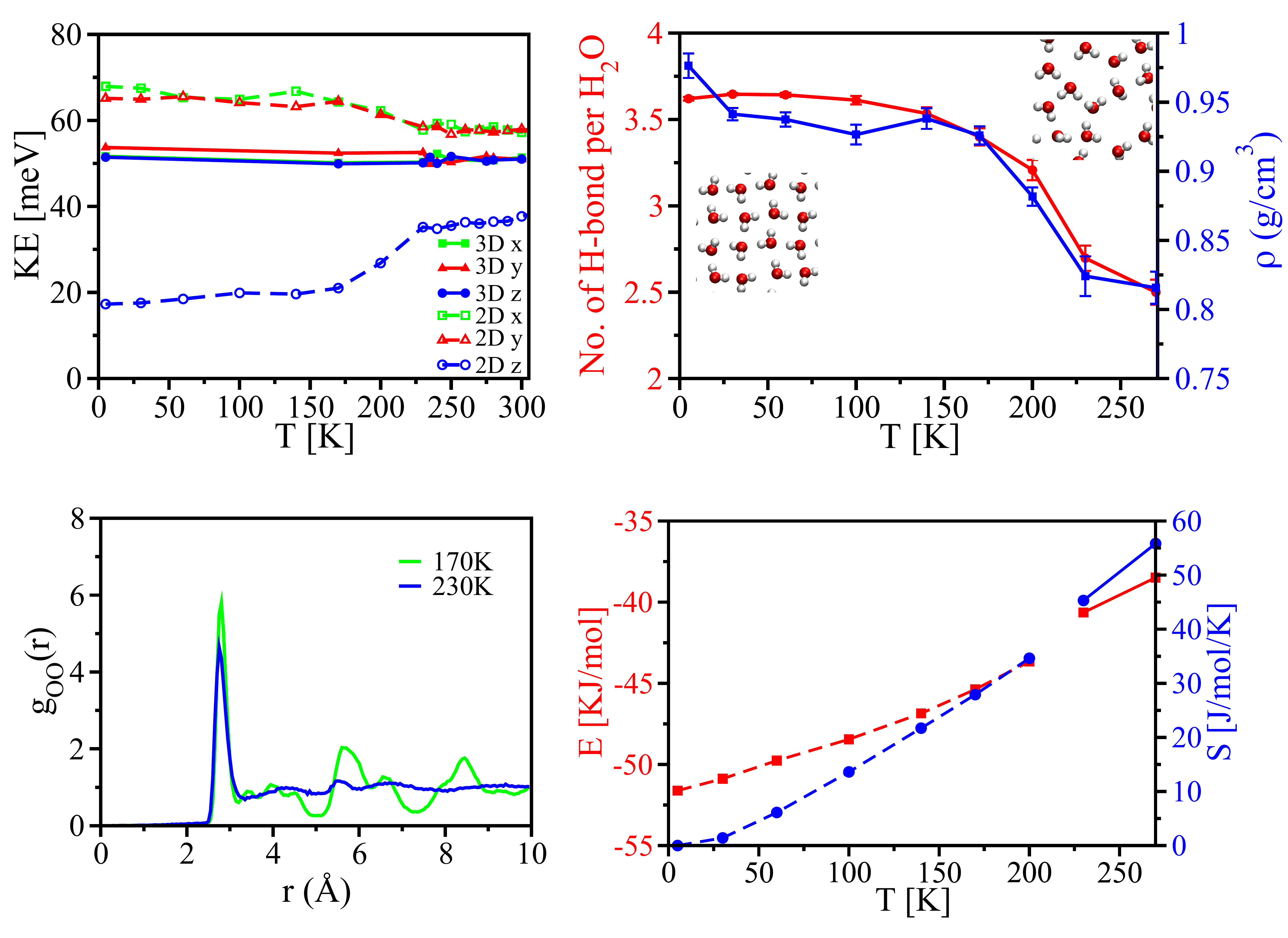}
\caption{(a) Temperature dependence of the directional component of Ke(H) for the 2D and bulk cases. (b) Number of hydrogen bond per water molecules (red) and density of water inside the nanopore (blue) are constant up to 170 K and decreases rapidly above that temperature. (c) Oxygen-Oxygen 2D radial distribution function of water confined inside graphene nanoslit. (d) Entropy (blue) and enthalpy (red) with varying temperature reveals discontinuity around 200 K.}
\label{KE_T}
\end{center}
\end{figure}

The upper panel of  Fig. \ref{DoS} depicts a total (spatial averaged) H-VDOS of an arbitrary H atom at 5 K, illustrating the typical pVDOS spectra used to deduce the temperature dependence of Ke(H) in bulk ice. Also shown in the lower panels of Fig. \ref{DoS} are the directional VDOS profiles,  computed for the H1 proton in a single H$_2$O in bulk ice at 5 K.
The VDOS profiles were calculated for all water molecules composing the 3D and 2D model systems and then utilized to deduce the average total and directional Ke(H) values versus temperature. The temperature dependence of the directional Ke(H) values is presented in Fig. \ref{KE_T}b.\par
We first note in the 3D bulk  (solid lines in Fig. \ref{KE_T}a) all directional Ke(H) components practically have the same value irrespective of the temperature, which well accords with  the fact that on average, in bulk water phases, the H$_2$O molecules are isotropically oriented. In clear contrast, however, for the 2D confined water monolayer (dashed lines in Fig. 3a), marked differences are noted between the Ke(H) components along the (z) direction and perpendicular to it (x,y). Moreover, the two "in-plane" monolayer components, Ke$_x$(H) and Ke$_y$(H), practically show the same values irrespective of the temperature due to the isotropy of the H$_2$O planes along the (x,y) plane, whereas those along the confining ("out-of-plane") direction, Ke$_z$(H), show far much lower values. Also, the in-plane and out-of-plane Ke(H) components reveal an anti correlated, or mirror image behavior [Fig. \ref{KE_T}(a)]. The isotropy, which leads to similar Ke(H) values along the X and Y directions, can be easily understood if one considers that the H$_2$O molecules composing the 2D monolayer, lay rather flat with respect to the graphene surface and on average, with no preferable in-plane orientation.\par 
\begin{table*}
\centering
\caption{Calculated and measured Ke(H) components in the molecular frame of reference of water under 3D (bulk) and 2D (H$_2$O@GOs) topologies. The ax, prp, mp directions with respect to the water molecule are defined in Fig. \ref{system} and accord with those of Refs. \cite{Finkelstein2019} and \cite{Romanelli2015}. Ke$_{tot}$(H) is the algebraic sum over the three directional components. PIMD, INS and SE stand for the path integral molecular dynamics calculation, inelastic neutron scattering, and the semi empirical approach respectively. The SE values of the 271 solid 293 K liquid were calculated using the formulation presented in the supporting information of Ref. \cite{Parmentier2015} and by utilizing the structural and spectroscopic parameters of liquid water.}
\label{table}
\begin{center}
\resizebox{\textwidth}{!}{\begin{tabular}{|c|cccccccccc|}
\hline
&\multicolumn{8}{c|}{3D}&\multicolumn{2}{c|}{2D}\\
\cline{2-11}
&\multicolumn{6}{c|}{Ice}&\multicolumn{2}{c|}{Liquid}&\multicolumn{2}{c|}{H$_2$O@GOs}\\
\cline{2-11}
Ke(H)&\multicolumn{6}{c|}{271K}&\multicolumn{1}{c|}{285K}&\multicolumn{1}{c|}{293K}&\multicolumn{1}{c|}{20K}&\multicolumn{1}{c|}{293K}\\
\cline{2-11}
&\multicolumn{4}{c|}{Calc.}&\multicolumn{2}{c|}{Exp.}&\multicolumn{1}{c|}{Exp.}&\multicolumn{1}{c|}{Calc.}&\multicolumn{2}{c|}{Exp.}\\
\cline{2-11}
&\multicolumn{1}{c|}{DFT\footnote{Ref. \cite{Finkelstein2019}}}&\multicolumn{2}{c|}{SE}&\multicolumn{1}{c|}{PIMD\footnote{Ref. \cite{Finkelstein1999,kapil2018}}}&\multicolumn{1}{c|}{DINS\footnote{Ref. \cite{Andreani2016}}}&\multicolumn{1}{c|}{DINS\footnote{Ref. \cite{Flammini2012}}}&\multicolumn{1}{c|}{DINS\footnote{Ref. \cite{Flammini2012}}}&\multicolumn{1}{c|}{SE}&\multicolumn{2}{c|}{DINS\footnote{Ref. \cite{Romanelli2015}}}\\
\cline{3-4}
&\multicolumn{1}{c|}{}&\multicolumn{1}{c|}{INS\footnote{Ref. \cite{Andreani2013}}}&\multicolumn{1}{c|}{Present}&\multicolumn{1}{c|}{}&\multicolumn{1}{c|}{}&\multicolumn{1}{c|}{}&\multicolumn{1}{c|}{}&\multicolumn{1}{c|}{}&\multicolumn{2}{c|}{}\\
\hline
prp&\multicolumn{1}{c|}{30.9}&\multicolumn{1}{c|}{21.6 $\pm$ 0.3}&\multicolumn{1}{c|}{23 $\pm$ 3}&\multicolumn{1}{c|}{24.2}&\multicolumn{1}{c|}{28 $\pm$ 2}&\multicolumn{1}{c|}{28.9 $\pm$ 2}&\multicolumn{1}{c|}{18.3 $\pm$ 2}&\multicolumn{1}{c|}{17.7 $\pm$ 3}&\multicolumn{1}{c|}{26 $\pm$ 7}&\multicolumn{1}{c|}{24 $\pm$ 6}\\
\hline
mp&\multicolumn{1}{c|}{37.4}&\multicolumn{1}{c|}{34.4 $\pm$ 1.2}&\multicolumn{1}{c|}{36 $\pm$ 3}&\multicolumn{1}{c|}{31.8}&\multicolumn{1}{c|}{38 $\pm$ 5}&\multicolumn{1}{c|}{38.1 $\pm$ 3}&\multicolumn{1}{c|}{51.8 $\pm$ 4}&\multicolumn{1}{c|}{38.9 $\pm$ 3}&\multicolumn{1}{c|}{45 $\pm$ 9}&\multicolumn{1}{c|}{45 $\pm$ 9}\\
\hline
ax&\multicolumn{1}{c|}{87.7}&\multicolumn{1}{c|}{98.8 $\pm$ 1.2}&\multicolumn{1}{c|}{93 $\pm$ 3}&\multicolumn{1}{c|}{99.5}&\multicolumn{1}{c|}{91 $\pm$ 5}&\multicolumn{1}{c|}{86.7 $\pm$ 3}&\multicolumn{1}{c|}{83.8 $\pm$ 3}&\multicolumn{1}{c|}{99.8 $\pm$ 3}&\multicolumn{1}{c|}{84 $\pm$ 12}&\multicolumn{1}{c|}{86 $\pm$ 12}\\
\hline
Ke$_{tot}$(H)&\multicolumn{1}{c|}{156}&\multicolumn{1}{c|}{154 $\pm$ 2}&\multicolumn{1}{c|}{152 $\pm$ 5}&\multicolumn{1}{c|}{155.5}&\multicolumn{1}{c|}{157 $\pm$ 2}&\multicolumn{1}{c|}{153.7 $\pm$ 2}&\multicolumn{1}{c|}{153.9 $\pm$ 3}&\multicolumn{1}{c|}{156.4 $\pm$ 5}&\multicolumn{1}{c|}{154.9 $\pm$ 2}&\multicolumn{1}{c|}{156.1 $\pm$ 2}\\
\hline
\end{tabular}}%
    
\end{center}
\end{table*}
It is also interesting to note that the 2D case reveals a phase transition occurring across the $\sim$ [170K-230K] temperature range. This is nicely reflected also by the instantaneous snapshots of the system at 170 K and 230 K [Fig. \ref{KE_T}(b)] that illustrate a order-disorder transition.
To characterise this transition, we computed the entropy and enthalpy\cite{lin2003two,lin2010two} of the confined water, both characterized by a clear jump at 200 K. These jumps reveal a first order phase transition of the water molecules  inside the graphene nanopore\cite{kumar2018}. Furthermore, the calculated water density inside the graphene nanopore and number of HBs per H$_2$O molecule versus temperature [Fig. \ref{KE_T}(b)], are also both in accordance with a order-disorder phase transition around 200 K. While the number of HBs per H$_2$O is constant up to 170 K, it decreases rather sharply above that temperature. In order to quantify the structural changes across the transition, the calculation was supplemented with that of the radial distribution function at 170 and 230 K [Fig. \ref{KE_T}(c)], showing a loss of ordering above 170 K. All in all,  we quantified this transition as a first-order phase transition.\par
\begin{figure}[!ht]
\begin{center}
\includegraphics[width=3.375in,keepaspectratio=true]{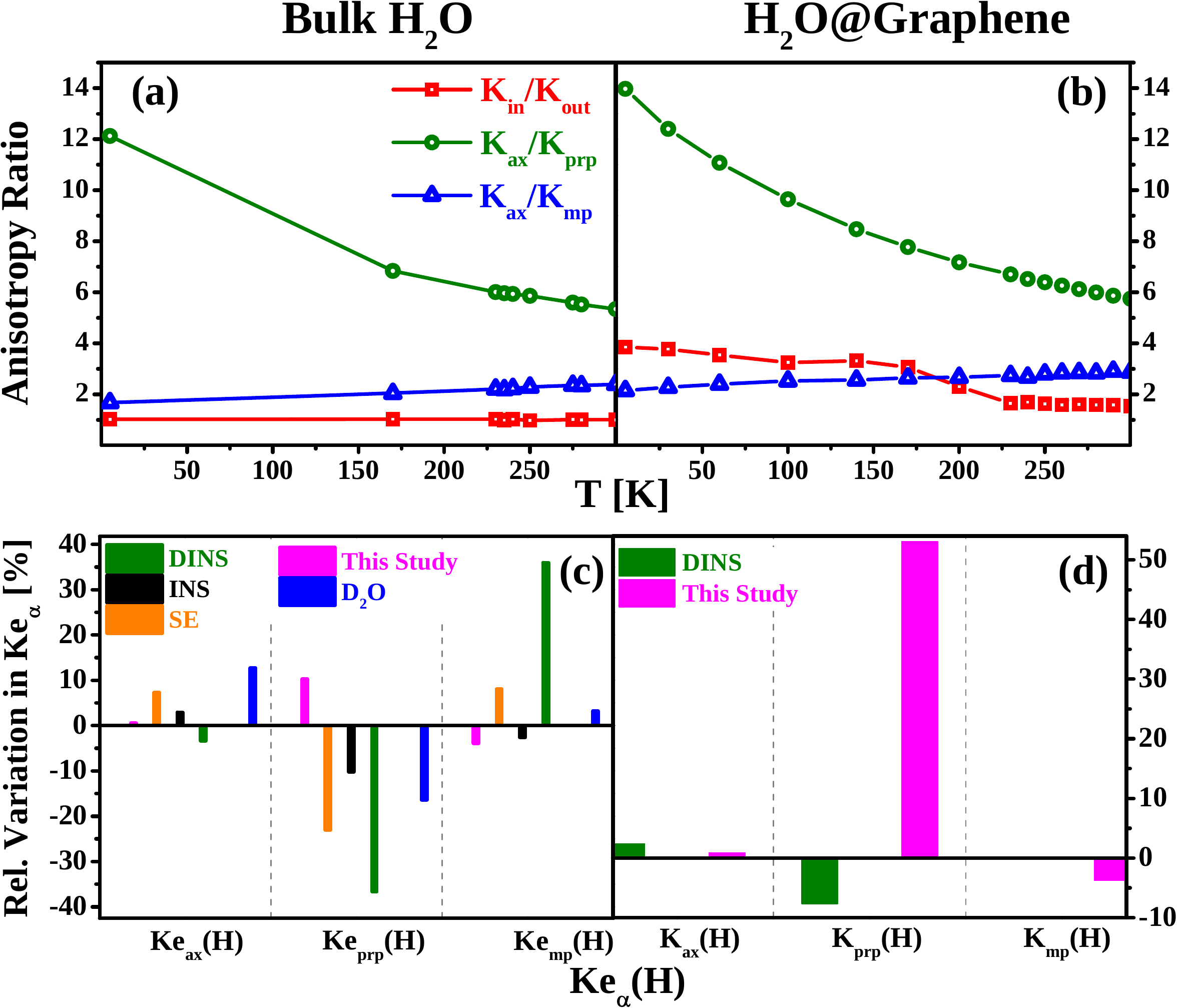}
\caption{Anisotropy ratios of Ke(H) versus T for the 3D bulk (a) and the confined 2D monolayer (b), in the molecular (circles and triangles) and Cartesian (squares) frames of reference. The percentage variation of Ke(H) components across phase transitions occurring in bulk (c) and confined (d) water, as measured by DINS (green) and INS (black) and calculated by SE (orange) and in the present study (red). Left (c) and right (d) Y-axes values correspond to a relative increase and decrease in the Ke(H) component respectively. For bulk water, the values correspond to temperatures increasing from 250 to 300K (present calculation), 271 to 285K (INS and DINS) and 271 to 293K (SE). Similarly, for the confined case, the involved temperatures are: 140 to 230K (present study) and 20 to 293K (DINS). Blue bars correspond to DINS results of bulk D$_2$O (see text).\cite{Romanelli2013}}
\label{Aniso}
\end{center}
\end{figure}

It is interesting to note that this transition is also captured by the variation in the kinetic energy tensor which changes upon confinement, depending on the order of confinement as well as on the dimension and nature of the confining geometry.\cite{Chakraborty2017,kumar2018,de2012,xu2011} The alteration in the kinetic energy tensor reflects a changed in the proton dynamics of the confined H$_2$O. In particular, it is seen in Fig. \ref{KE_T}(a) that while the directional components of Ke(H) in the case of the 3D bulk possess fairly equal magnitudes and monotonically evolve with temperature, those in the 2D case noticeably differ in magnitude and are accompanied by step like abruptions across the phase transition temperature.
The findings of  Fig. \ref{KE_T} hint for  an insightful way to capture the dynamics, namely to look at the evolution of the Ke(H) anisotropy versus temperature in both Cartesian and molecular frames of reference. In doing so we refer to three Ke(H) anisotropy ratios. The first one, in the Cartesian frame of reference, is of more relevance to the case of the lower dimensionality, and is defined [see Fig. \ref{Aniso}] as the ratio between the average in-plane, K$_a$ (H)=(Ke$_x$ (H)+Ke$_y$ (H))/2, and out of plane, K$_c$ (H)=Ke$_z$ (H), components. It evaluates the ratio between the average projections of Ke(H) along and perpendicular to the confining plane (z = constant). Hence, for the monolayer water that lays parallel to the graphene planes, a noticeable anisotropy may be expected, and moreover, any significant modification of the anisotropy ratio may suggest the occurrence of a structural reordering in the monolayer. In contrast, for the 3D case one may expect a practical isotropy, namely a unity ratio. The above is nicely evident by the red data points in Fig. \ref{Aniso} where K$_a$(H)/K$_c$(H) is nearly unity across the whole temperature range for the 3D case [Fig \ref{Aniso}(a)], while exhibiting a rich temperature dependence in the confined case [Fig. \ref{Aniso}(b)]. The temperature effect on the anisotropy ratio in the 2D case [red square in Fig. \ref{Aniso}(b)] is "commenced" at 5K with a rather large value of $\sim$3.9, moderately decreasing to $\sim$3.1 at 170K, then steeply declines down to $\sim$1.7 at 230K, hence nicely accords with the above discussed phase transition, and again continuously, but moderately, falls down to $\sim$1.5 at room temperature (RT).\par
The other two anisotropy ratios [circles and triangles in Fig. \ref{Aniso}] are defined in the molecular frame of reference [see Fig. \ref{system}(c)] which provides a means to look for signatures of competing quantum effects in Ke(H) such as those reported by DINS to occur upon melting of heavy water.\cite{Romanelli2013} The first ratio is defined between K$_{ax}$(H), the component of Ke(H) along the OH bond, and K$_{prp}$(H), that along the normal to the H$_2$O plane. Similarly, the remaining anisotropy ratio is defined as that between K$_{ax}$(H) and K$_{mp}$(H), the component of Ke(H) that lays in the H$_2$O plane but directed normal to K$_{ax}$(H). A marked but rather similar K$_{ax}$(H)/K$_{prp}$(H) anisotropy [green curves in Figs. \ref{Aniso}(a) and (b)] is noticed for the 3D and 2D cases, which accords with the fact that up to 300K the molecular structure is after all preserved regardless of the water phase.
This ratio depicts in fact the difference between the OH bond strength and that corresponding to the motions of the proton out of the H$_2$O molecular plane. In either of the 2D and 3D cases, the average number of HBs per water molecule decreases with increasing temperature [Fig. \ref{KE_T}(b)], increasing in turn the OH bond strength. At the same time, however, the translational and librational motions are gradually diminished due to increased out of plane thermal motions, therefore the anisotropy is overall depressed.
Since the average number of HBs at any given temperature is smaller in the 2D water monolayer, the OH bonds are strengthened compared to the 3D case, which increases the K$_{ax}$ of 2D confined water compared to the bulk phase.
As a result, the K$_{ax}$(H)/K$_{prp}$(H) anisotropy in the molecular frame of reference in the 2D case is relatively higher compared to the 3D bulk [Fig \ref{Aniso}].
The calculated values of the anisotropy ratios versus temperature are shown in Fig. \ref{Aniso}.\par
Finally, the K$_{ax}$(H)/K$_{mp}$(H) ratio (blue curves in Fig. \ref{Aniso}) increases between 5 K and 300 K as a gradual and monotonous function, from $\sim$1.7 to $\sim$2.2 and from $\sim$2.1 to $\sim$2.9, for the 2D [Fig. \ref{Aniso}(b)] and bulk [Fig. \ref{Aniso}(a)] cases respectively. Interestingly, the K$_{ax}$(H)/K$_{mp}$(H) curve of the 2D system intersects that of K$_{ax}$(H)/K$_{prp}$(H) at about the median temperature of the transition range [Fig. \ref{Aniso}(b)] across which the phase transition is detected.\par
It would be interesting to validate the findings of Fig \ref{Aniso} against experiment, namely DINS results. Unfortunately the DINS reports on water phases provide the directional Ke(H) values only in the molecular frame of reference. Nonetheless, an indirect experimental evidence for the above may be obtained by accounting for a very recent DINS study which was focused on investigating the effective isotropy of the proton local potential in a biphenyl hydrocarbon.\cite{Finkelstein1999} As in biphenyl, here too, the anisotropy is gradually diminished as the number of water units increase on going from the confined case of a water monolayer to the 3D bulk. This scenario is clearly reflected in Fig \ref{Aniso} for the molecular frame of reference, where the anisotropy ratio is insensitive in the 3D case [Fig. \ref{Aniso}(a)] to the solid liquid transition, while revealing a clear sensitivity in the 2D case for the order-disorder transition [Fig. \ref{Aniso}(b)]. \par
Lastly, in former DINS studies the relative values of the directional Ke(H) components in the molecular frame of reference, as assumed by the protons across phase transitions of 2D and bulk water, were understood as signatures of competing quantum effects (CQEs) on Ke(H) upon the transition. It is thus of interest to examine also here how the directional components of Ke(H) in the molecular frame of reference vary across the phase transition in view of those reported by DINS for bulk water and water@Graphene. In the following discussion we also refer to the values presented in Table \ref{table} which include SE calculations using INS data\cite{Andreani2017}. The data points in Fig. \ref{Aniso} depict the percentage by which the Ke(H) components, for the molecular frame of reference in bulk water and in water@Graphene, vary across phase transition temperatures. \par

By reviewing the values of Figs. \ref{Aniso}(c) and \ref{Aniso}(d), no detectable distinction is noticed between the 3D and 2D systems, however the following three common features may be noted for the two cases. First, K$_{prp}$(H), the out of plane component of Ke(H), and K$_{ax}$(H), along the OH bond, seem to undergo the largest and smallest relative variations respectively across the phase transition. Second, in all cases at least one of the three components varies in opposite direction compared to the remaining two components. Lastly, the overall effect of these variations is of a compensating nature, namely such that impose a very small change in the total Ke(H) across the phase transition, hence the total Ke(H) reveals an overall picture of the same dynamics for the cases of the bulk and confined water. These particular features were previously observed by DINS measurements on bulk D$_2$O across the melting point.\cite{Romanelli2013} The directional components of the kinetic energy of the deutron, in the molecular frame of reference, were extracted in Ref. \cite{Romanelli2013} (as in Refs. \cite{Romanelli2015,Flammini2012,Andreani2013,Andreani2016}) from the experimental momentum distribution by accounting for its measured profile as arising from the spherical average of an anisotropic Gaussian distribution. The [K$_{ax}$(D), K$_{prp}$(D), K$_{mp}$(D)] values (in meV units) were [48.1$\pm$3.4, 22.5$\pm$1.8, 37.4$\pm$2.5] and [54.2$\pm$2.4, 18.8$\pm$1.1, 38.6$\pm$2.5] at 274K (solid) and 280K (liquid) respectively. In comparison to Fig. \ref{Aniso}(c), namely in terms of the percentage change going from the solid to the liquid phase in bulk H$_2$O, K$_{ax}$(D), K$_{prp}$(D) and K$_{mp}$(D) vary by $+$12.7\%, -16.4\% and +3.2\% respectively [denoted by blue bars in Fig. \ref{Aniso}(c)], with a corresponding total Ke(D) increment of ca. 3.3\%, from 108.0$\pm$2.0 to 111.6$\pm$2.0 meV. This generic behaviour is generally used to assign the CQEs on Ke(H) upon melting,\cite{Romanelli2013} a behaviour which also has been observed in the present calculation.\par
In the present study, we explored the directional components of the ke(H) tensor for the cases of bulk water and a 2D water monolayer confined inside a graphene slit pore. We show that  the flexible TIP4P-2005 model provides general good agreement with DINS measured values for bulk Ice-Ih and liquid over the entire studied temperature range between 5K and RT. More importantly, the calculated temperature effect on the directional Ke(H) tensor in water clearly reveals  a significant change near 200 K which is indicative of a order-disorder phase transition. This transition  is also reflected in the water density and average number of HBs per water molecule. \par
Interestingly, we also found that only a particular anisotropy ratio, namely that in the \textit{Cartesian} frame of reference, acquires a sharp jump near the transition temperature, while that in the \textit{molecular} frame of reference remains practically intact, that is, insensitive to the phase transition. Hence, we can conclude that depending on the geometry of confinement, Ke(H) anisotropy ratios may be used as a valuable choice of an order parameter for detecting order-disorder phase transitions.\par
\section{\large ACKNOWLEDGMENT}
We acknowledge DST, India, for the computational support through TUECMS machine, IISc. We thank the DAE, India, and DST, India, for financial support. M.M. acknowledges the Councel of Sceintific and Industrial Research (CSIR), India, for financial
support.
\section{Appendix}
\subsection{1. Theoretical Framework}
To simulate the directional components of Ke(H) we start by constructing the kinetic energy tensor, whose principal components determine the kinetic energy values along the orthogonal direction of that coordinate system.
The kinetic energy tensor in the Cartesian coordinate system may be expressed as:\par
\begin{center}
	$$
\begin{pmatrix}
Ke_{xx} & Ke_{xy} & Ke_{xz}\\
Ke_{yx} & Ke_{yy} & Ke_{yz}\\
Ke_{zx} & Ke_{zy} & Ke_{zz}\\
\end{pmatrix}
        $$
\end{center}
$Ke_{ij}$ denotes the atomic kinetic energy along that direction, with i and j standing for x,y or z direction.\par
Atomic kinetic energies in various phases can be calculated using different methods as discussed in detail in the literature.\cite{Finkelstein2019,Parmentier2015} Here, Ke(H) in condensed water phases is computed from the simulated pVDOS of the system, g$^{H}_{ij}(n)$,
i.e. the average projection of the total VDOS of the protons, by taking each vibrational state as a \textit{quantum harmonic} oscillator using following equation:
\begin{equation}
    Ke_{ij}(H) = \frac{\frac{3}{2} \int_{\nu_{\scaleobj{.6}0}}^{\nu_{\scaleobj{.6}f}} \text{g}^{H}_{ij}(\nu) \alpha(\nu,T) d\nu}{\int_{\nu_{\scaleobj{.6}0}}^{\nu_{\scaleobj{.6}f}} \text{g}^{H}_{ij}(\nu)d\nu} 
\end{equation}
where, g$_{ij}^{H}$($\nu$) is the partial vibrational density of state (VDOS) of H atom with $\nu_{0}$ and $\nu_{f}$ frequency limits, and $\alpha$($\nu$,$T$) is the total energy of the quantum harmonic oscillator of frequency, $\nu$, at temperature $T$ given by\par
\begin{equation}
        \alpha (\nu,T) = h\nu \left(\frac{1}{exp(\beta h\nu) - 1} + \frac{1}{2}\right),
\end{equation}
        where, $\beta = \frac{1}{k_{B}T}$ and \textit{h} is plank constant.\par
        The partial VDOS, \text{g}$_{ij}^{H}$($\nu$), (H-VDOS of atom H) is calculated in MD simulation using the mass weighted sum of atomic spectral densities:
\begin{equation}
        \text{g}_{ij}^{H}(\nu) = \frac{2}{KT} \sum_{j=1}^{N}\sum_{k=1}^{3} m_{k} s_{ij}^{k}(\nu) ,
\end{equation}
Where, m$_k$, is the atomic mass, k stands for the atom index for $k^{th}$ hydrogen in the system (e.g., 2N hydrogen atoms in a system comprising a total number of N water molecules). s$_{ij}^{k}$($\nu$) is defined as the spectral density of the k$^{th}$ H-atom, and is computed from the square of the Fourier transform of the corresponding velocities:
\begin{equation}
        s_{ij}^{k}(\nu) = \lim_{\tau\to\infty} \frac{1}{2\tau} {\int_{-\tau}^{\tau}\int_{-\tau}^{\tau}V_{i}^{k}(t).V_{j}^{k}(t+t^{'})dt^{'}} e^{-\iota2\pi \nu t}dt,
\end{equation}
Where, V$_{i}^{k}$ is the velocity of the $k^{th}$ H-atom along the $i^{th}$ direction at time $t$. The momentum-momentum correlation in two different directions does not correlate in a classical picture hence the off-diagonal term of the kinetic energy matrix becomes zero. Accordingly, the tensor can be rewritten in the form of a diagonalized tensor as follows:
\begin{center}
	$$
\begin{pmatrix}
Ke_{x}  & 0       & 0\\
0       & Ke_{y}  & 0\\
0       & 0       & Ke_{z}\\
\end{pmatrix}
         $$
\end{center}
$Ke_{i}$ the kinetic energy value along the i = x,y,z direction. Note that the spectral density of atoms can also be deduced from the velocity-velocity autocorrelation. One can write the total Ke(H) by summing over  the directional components:
\begin{equation}
       Ke(H) = Ke_x(H) + Ke_y(H) + Ke_z(H),
\end{equation}
Integrating the VDOS over the whole frequency range will provide the total number of degrees of freedom of the system in the i$^{th}$ direction:
One can write the total Ke(H) by summing over  the directional components:
\begin{equation} \int_{0}^{\infty} g_x(\nu)d(\nu)=3N \end{equation}

\subsection{2. Molecular Simulation details}
We have performed MD simulation for hexagonal ice of 432 water molecules (3D case) and monolayer of water confined inside the graphene nano slit pore separated by 6.8 {\AA} (2D case). For 3D water, we initiated the simulation at 5 K and 1 bar pressure with the initial hexagonal ice structures in the NPT ensemble. After 50 ns long NPT simulation, the final configuration from 5 K run was used as input to the next higher temperature of 100 K and so on. The simulations were performed at 100, 140, 170, 230, 240, 250, 270, 280, 300 K. NVT simulation was performed for 50 ns for each of the above mentioned temperature using the equilibrated configuration obtained in the NPT ensemble. The velocity-rescale\cite{Bussi2007} thermostat and Parrinello-Rahman barostat were used in NPT simulation with a temperature coupling constant of 1.3 ps and pressure coupling constant of 2 ps with the integration time step of 1 fs. Periodic boundary condition was used in all the three directions.\par
For 2D case, we initiated the simulation at 5 K and 1 bar pressure with the initial square ice structures between two graphene sheet separated by 6.8 {\AA}, embedded in a 70 $\times$ 70 $\times$ 37 {\AA}$^{3}$ water bath. We performed 40 ns NPT simulation at 5, 30, 60, 100, 140, 170, 200, 230, 240, 250, 260, 270, 280, 290 and 300 K at 1 atm pressure in the NPT ensemble. After 40 ns equilibrium simulation in the NPT ensemble, we performed 10ns long NVT run. Throughout the simulation, graphene sheets were held fixed to its initial position using a harmonic restraint using a force constant of 1000 KJ/mol/nm$^{2}$ in each direction. We used Visual Molecular Dynamics (VMD) software\cite{Humphrey1996} to build graphene sheet of 4 $\times$ 4 nm$^2$ and xLeap module of AMBER\cite{Case2020} to solvate the nanopore using the TIP4P-2005 flexible water model.\cite{Abascal2005,Gonzalez2011}\par
In each case, After NVT simulation, we performed another set of NVT simulation for 500 ps and save trajectory in every 2 femtosecond to compute density of state using velocity autocorrelation. A similar protocol was used in our earlier simulations of bulk and confined water.\cite{Moid2021,Raghav2015,Majumdar2021} The directional component of Ke(H) is averaged over at least 10 molecules for high temperature ($>$200 K) and up to 100 molecules at low temperature ($<$200 K).

\bibliographystyle{ieeetr}
\bibliography{references}

\begin{thebibliography}{10}

\bibitem{David2019}
R.~O. David, C.~Marcolli, J.~Fahrni, Y.~Qiu, Y.~A.~P. Sirkin, V.~Molinero,
  F.~Mahrt, D.~Br{\"u}hwiler, U.~Lohmann, and Z.~A. Kanji, ``Pore condensation
  and freezing is responsible for ice formation below water saturation for
  porous particles,'' {\em Proceedings of the National Academy of Sciences},
  vol.~116, no.~17, pp.~8184--8189, 2019.

\bibitem{Mochizuki2015}
K.~Mochizuki and K.~Koga, ``Solid- liquid critical behavior of water in
  nanopores,'' {\em Proceedings of the National Academy of Sciences}, vol.~112,
  no.~27, pp.~8221--8226, 2015.

\bibitem{Koga2001}
K.~Koga, G.~Gao, H.~Tanaka, and X.~C. Zeng, ``Formation of ordered ice
  nanotubes inside carbon nanotubes,'' {\em Nature}, vol.~412, no.~6849,
  pp.~802--805, 2001.

\bibitem{Algara2015}
G.~Algara-Siller, O.~Lehtinen, F.~Wang, R.~R. Nair, U.~Kaiser, H.~Wu, A.~K.
  Geim, and I.~V. Grigorieva, ``Square ice in graphene nanocapillaries,'' {\em
  Nature}, vol.~519, no.~7544, pp.~443--445, 2015.

\bibitem{Holt2006}
J.~K. Holt, H.~G. Park, Y.~Wang, M.~Stadermann, A.~B. Artyukhin, C.~P.
  Grigoropoulos, A.~Noy, and O.~Bakajin, ``Fast mass transport through
  sub-2-nanometer carbon nanotubes,'' {\em Science}, vol.~312, no.~5776,
  pp.~1034--1037, 2006.

\bibitem{Tunuguntla2017}
R.~H. Tunuguntla, R.~Y. Henley, Y.-C. Yao, T.~A. Pham, M.~Wanunu, and A.~Noy,
  ``Enhanced water permeability and tunable ion selectivity in subnanometer
  carbon nanotube porins,'' {\em Science}, vol.~357, no.~6353, pp.~792--796,
  2017.

\bibitem{Secchi2016}
E.~Secchi, S.~Marbach, A.~Nigu{\`e}s, D.~Stein, A.~Siria, and L.~Bocquet,
  ``Massive radius-dependent flow slippage in carbon nanotubes,'' {\em Nature},
  vol.~537, no.~7619, pp.~210--213, 2016.

\bibitem{Mukherjee2007}
B.~Mukherjee, P.~K. Maiti, C.~Dasgupta, and A.~Sood, ``Strong correlations and
  fickian water diffusion in narrow carbon nanotubes,'' {\em The Journal of
  chemical physics}, vol.~126, no.~12, p.~124704, 2007.

\bibitem{Bonthuis2011}
D.~J. Bonthuis, K.~F. Rinne, K.~Falk, C.~N. Kaplan, D.~Horinek, A.~N. Berker,
  L.~Bocquet, and R.~R. Netz, ``Theory and simulations of water flow through
  carbon nanotubes: prospects and pitfalls,'' {\em Journal of Physics:
  Condensed Matter}, vol.~23, no.~18, p.~184110, 2011.

\bibitem{Naskar2020}
S.~Naskar, A.~K. Sahoo, M.~Moid, and P.~K. Maiti, ``Ultra-high permeable
  phenine nanotube membranes for water desalination,'' {\em arXiv preprint
  arXiv:2005.12145}, 2020.

\bibitem{Noon2002}
W.~H. Noon, K.~D. Ausman, R.~E. Smalley, and J.~Ma, ``Helical ice-sheets inside
  carbon nanotubes in the physiological condition,'' {\em Chemical Physics
  Letters}, vol.~355, no.~5-6, pp.~445--448, 2002.

\bibitem{Bai2006}
J.~Bai, J.~Wang, and X.~C. Zeng, ``Multiwalled ice helixes and ice nanotubes,''
  {\em Proceedings of the National Academy of Sciences}, vol.~103, no.~52,
  pp.~19664--19667, 2006.

\bibitem{Nomura2017}
K.~Nomura, T.~Kaneko, J.~Bai, J.~S. Francisco, K.~Yasuoka, and X.~C. Zeng,
  ``Evidence of low-density and high-density liquid phases and isochore end
  point for water confined to carbon nanotube,'' {\em Proceedings of the
  National Academy of Sciences}, vol.~114, no.~16, pp.~4066--4071, 2017.

\bibitem{Majumdar2021}
J.~Majumdar, M.~Moid, C.~Dasgupta, and P.~K. Maiti, ``Dielectric profile and
  electromelting of a monolayer of water confined in graphene slit pore,'' {\em
  The Journal of Physical Chemistry B}, 2021.

\bibitem{Joseph2008}
S.~Joseph and N.~Aluru, ``Why are carbon nanotubes fast transporters of
  water?,'' {\em Nano letters}, vol.~8, no.~2, pp.~452--458, 2008.

\bibitem{Shiomi2007}
J.~Shiomi, T.~Kimura, and S.~Maruyama, ``Molecular dynamics of ice-nanotube
  formation inside carbon nanotubes,'' {\em The Journal of Physical Chemistry
  C}, vol.~111, no.~33, pp.~12188--12193, 2007.

\bibitem{Chakraborty2017}
S.~Chakraborty, H.~Kumar, C.~Dasgupta, and P.~K. Maiti, ``Confined water:
  structure, dynamics, and thermodynamics,'' {\em Accounts of chemical
  research}, vol.~50, no.~9, pp.~2139--2146, 2017.

\bibitem{Mukherjee2008}
B.~Mukherjee, P.~K. Maiti, C.~Dasgupta, and A.~K. Sood, ``Strongly anisotropic
  orientational relaxation of water molecules in narrow carbon nanotubes and
  nanorings,'' {\em Acs Nano}, vol.~2, no.~6, pp.~1189--1196, 2008.

\bibitem{Schlaich2016}
A.~Schlaich, E.~W. Knapp, and R.~R. Netz, ``Water dielectric effects in planar
  confinement,'' {\em Physical review letters}, vol.~117, no.~4, p.~048001,
  2016.

\bibitem{kayal2015}
A.~Kayal and A.~Chandra, ``Exploring the structure and dynamics of
  nano-confined water molecules using molecular dynamics simulations,'' {\em
  Molecular Simulation}, vol.~41, no.~5-6, pp.~463--470, 2015.

\bibitem{Lupi2014}
L.~Lupi and V.~Molinero, ``Does hydrophilicity of carbon particles improve
  their ice nucleation ability?,'' {\em The Journal of Physical Chemistry A},
  vol.~118, no.~35, pp.~7330--7337, 2014.

\bibitem{Whitby2008}
M.~Whitby, L.~Cagnon, M.~Thanou, and N.~Quirke, ``Enhanced fluid flow through
  nanoscale carbon pipes,'' {\em Nano letters}, vol.~8, no.~9, pp.~2632--2637,
  2008.

\bibitem{Falk2010}
K.~Falk, F.~Sedlmeier, L.~Joly, R.~R. Netz, and L.~Bocquet, ``Molecular origin
  of fast water transport in carbon nanotube membranes: superlubricity versus
  curvature dependent friction,'' {\em Nano letters}, vol.~10, no.~10,
  pp.~4067--4073, 2010.

\bibitem{Reiter2012}
G.~F. Reiter, A.~I. Kolesnikov, S.~J. Paddison, P.~Platzman, A.~P. Moravsky,
  M.~A. Adams, and J.~Mayers, ``Evidence for an anomalous quantum state of
  protons in nanoconfined water,'' {\em Physical Review B}, vol.~85, no.~4,
  p.~045403, 2012.

\bibitem{Kolesnikov2016}
A.~I. Kolesnikov, G.~F. Reiter, N.~Choudhury, T.~R. Prisk, E.~Mamontov,
  A.~Podlesnyak, G.~Ehlers, A.~G. Seel, D.~J. Wesolowski, and L.~M. Anovitz,
  ``Quantum tunneling of water in beryl: a new state of the water molecule,''
  {\em Physical review letters}, vol.~116, no.~16, p.~167802, 2016.

\bibitem{Senesi2007}
R.~Senesi, A.~Pietropaolo, A.~Bocedi, S.~Pagnotta, and F.~Bruni, ``Proton
  momentum distribution in a protein hydration shell,'' {\em Physical review
  letters}, vol.~98, no.~13, p.~138102, 2007.

\bibitem{Pantalei2011}
C.~Pantalei, R.~Senesi, C.~Andreani, P.~Sozzani, A.~Comotti, S.~Bracco,
  M.~Beretta, P.~E. Sokol, and G.~Reiter, ``Interaction of single water
  molecules with silanols in mesoporous silica,'' {\em Physical Chemistry
  Chemical Physics}, vol.~13, no.~13, pp.~6022--6028, 2011.

\bibitem{Romanelli2015}
G.~Romanelli, R.~Senesi, X.~Zhang, K.~P. Loh, and C.~Andreani, ``Probing the
  effects of 2d confinement on hydrogen dynamics in water and ice adsorbed in
  graphene oxide sponges,'' {\em Physical Chemistry Chemical Physics}, vol.~17,
  no.~47, pp.~31680--31684, 2015.

\bibitem{Andreani2017}
C.~Andreani, R.~Senesi, M.~Krzystyniak, G.~Romanelli, and F.~Fernandez-Alonso,
  ``Atomic quantum dynamics in materials research,'' in {\em Experimental
  Methods in the Physical Sciences}, vol.~49, pp.~403--457, Elsevier, 2017.

\bibitem{Colognesi2011}
D.~Colognesi, ``Extraction of single-particle mean kinetic energy from
  macroscopic thermodynamic data,'' {\em Physica B: Condensed Matter},
  vol.~406, no.~14, pp.~2723--2730, 2011.

\bibitem{Finkelstein2014}
Y.~Finkelstein and R.~Moreh, ``Temperature dependence of the proton kinetic
  energy in water between 5 and 673 k,'' {\em Chemical Physics}, vol.~431,
  pp.~58--63, 2014.

\bibitem{Flammini2012}
D.~Flammini, A.~Pietropaolo, R.~Senesi, C.~Andreani, F.~McBride, A.~Hodgson,
  M.~A. Adams, L.~Lin, and R.~Car, ``Spherical momentum distribution of the
  protons in hexagonal ice from modeling of inelastic neutron scattering
  data,'' {\em The Journal of chemical physics}, vol.~136, no.~2, p.~024504,
  2012.

\bibitem{Andreani2013}
C.~Andreani, G.~Romanelli, and R.~Senesi, ``A combined ins and dins study of
  proton quantum dynamics of ice and water across the triple point and in the
  supercritical phase,'' {\em Chemical Physics}, vol.~427, pp.~106--110, 2013.

\bibitem{Senesi2013}
R.~Senesi, G.~Romanelli, M.~Adams, and C.~Andreani, ``Temperature dependence of
  the zero point kinetic energy in ice and water above room temperature,'' {\em
  Chemical Physics}, vol.~427, pp.~111--116, 2013.

\bibitem{Pietropaolo2009}
A.~Pietropaolo, R.~Senesi, C.~Andreani, and J.~Mayers, ``Quantum effects in
  water: proton kinetic energy maxima in stable and supercooled liquid,'' {\em
  Brazilian Journal of Physics}, vol.~39, pp.~318--321, 2009.

\bibitem{Andreani2016}
C.~Andreani, G.~Romanelli, and R.~Senesi, ``Direct measurements of quantum
  kinetic energy tensor in stable and metastable water near the triple point:
  An experimental benchmark,'' {\em The journal of physical chemistry letters},
  vol.~7, no.~12, pp.~2216--2220, 2016.

\bibitem{Moid2020}
M.~Moid, Y.~Finkelstein, R.~Moreh, and P.~K. Maiti, ``Microscopic study of
  proton kinetic energy anomaly for nanoconfined water,'' {\em The Journal of
  Physical Chemistry B}, vol.~124, no.~1, pp.~190--198, 2019.

\bibitem{Abascal2005}
J.~L. Abascal and C.~Vega, ``A general purpose model for the condensed phases
  of water: Tip4p/2005,'' {\em The Journal of chemical physics}, vol.~123,
  no.~23, p.~234505, 2005.

\bibitem{Gonzalez2011}
M.~A. Gonz{\'a}lez and J.~L. Abascal, ``A flexible model for water based on
  tip4p/2005,'' {\em The Journal of chemical physics}, vol.~135, no.~22,
  p.~224516, 2011.

\bibitem{Mario2015}
M.~S.~F. Mario, M.~Neek-Amal, and F.~Peeters, ``Aa-stacked bilayer square ice
  between graphene layers,'' {\em Physical Review B}, vol.~92, no.~24,
  p.~245428, 2015.

\bibitem{Pascal2019}
T.~A. Pascal, C.~P. Schwartz, K.~V. Lawler, and D.~Prendergast, ``The purported
  square ice in bilayer graphene is a nanoscale, monolayer object,'' {\em The
  Journal of chemical physics}, vol.~150, no.~23, p.~231101, 2019.

\bibitem{Finkelstein2019}
Y.~Finkelstein, R.~Moreh, F.~Bianchini, and P.~Vajeeston, ``Anisotropy of the
  proton kinetic energy in ice ih,'' {\em Surface Science}, vol.~679,
  pp.~174--179, 2019.

\bibitem{Parmentier2015}
A.~Parmentier, J.~Shephard, G.~Romanelli, R.~Senesi, C.~Salzmann, and
  C.~Andreani, ``Evolution of hydrogen dynamics in amorphous ice with
  density,'' {\em The journal of physical chemistry letters}, vol.~6, no.~11,
  pp.~2038--2042, 2015.

\bibitem{Finkelstein1999}
Y.~Finkelstein, R.~Moreh, and O.~Shahal, ``Out-of-plane orientation of
  multilayer n2 films adsorbed on grafoil at 20 k,'' {\em Surface science},
  vol.~437, no.~3, pp.~265--276, 1999.

\bibitem{kapil2018}
V.~Kapil, A.~Cuzzocrea, and M.~Ceriotti, ``Anisotropy of the proton momentum
  distribution in water,'' {\em The Journal of Physical Chemistry B}, vol.~122,
  no.~22, pp.~6048--6054, 2018.

\bibitem{lin2003two}
S.-T. Lin, M.~Blanco, and W.~A. Goddard~III, ``The two-phase model for
  calculating thermodynamic properties of liquids from molecular dynamics:
  Validation for the phase diagram of lennard-jones fluids,'' {\em The Journal
  of chemical physics}, vol.~119, no.~22, pp.~11792--11805, 2003.

\bibitem{lin2010two}
S.-T. Lin, P.~K. Maiti, and W.~A. Goddard~III, ``Two-phase thermodynamic model
  for efficient and accurate absolute entropy of water from molecular dynamics
  simulations,'' {\em The Journal of Physical Chemistry B}, vol.~114, no.~24,
  pp.~8191--8198, 2010.

\bibitem{kumar2018}
H.~Kumar, C.~Dasgupta, and P.~K. Maiti, ``Phase transition in monolayer water
  confined in janus nanopore,'' {\em Langmuir}, vol.~34, no.~40,
  pp.~12199--12205, 2018.

\bibitem{de2012}
E.~de~la Llave, V.~Molinero, and D.~A. Scherlis, ``Role of confinement and
  surface affinity on filling mechanisms and sorption hysteresis of water in
  nanopores,'' {\em The Journal of Physical Chemistry C}, vol.~116, no.~2,
  pp.~1833--1840, 2012.

\bibitem{xu2011}
L.~Xu and V.~Molinero, ``Is there a liquid--liquid transition in confined
  water?,'' {\em The Journal of Physical Chemistry B}, vol.~115, no.~48,
  pp.~14210--14216, 2011.

\bibitem{Romanelli2013}
G.~Romanelli, M.~Ceriotti, D.~E. Manolopoulos, C.~Pantalei, R.~Senesi, and
  C.~Andreani, ``Direct measurement of competing quantum effects on the kinetic
  energy of heavy water upon melting,'' {\em The Journal of Physical Chemistry
  Letters}, vol.~4, no.~19, pp.~3251--3256, 2013.

\bibitem{Bussi2007}
G.~Bussi, D.~Donadio, and M.~Parrinello, ``Canonical sampling through velocity
  rescaling,'' {\em The Journal of chemical physics}, vol.~126, no.~1,
  p.~014101, 2007.

\bibitem{Humphrey1996}
W.~Humphrey, A.~Dalke, and K.~Schulten, ``Vmd: visual molecular dynamics,''
  {\em Journal of molecular graphics}, vol.~14, no.~1, pp.~33--38, 1996.

\bibitem{Case2020}
D.~A. Case, K.~Belfon, I.~Ben-Shalom, S.~R. Brozell, D.~Cerutti, T.~Cheatham,
  V.~W.~D. Cruzeiro, T.~Darden, R.~E. Duke, G.~Giambasu, {\em et~al.}, ``Amber
  2020,'' 2020.

\bibitem{Moid2021}
M.~Moid, S.~Sastry, C.~Dasgupta, T.~A. Pascal, and P.~K. Maiti,
  ``Dimensionality dependence of the kauzmann temperature: A case study using
  bulk and confined water,'' {\em The Journal of Chemical Physics}, vol.~154,
  no.~16, p.~164510, 2021.

\bibitem{Raghav2015}
N.~Raghav, S.~Chakraborty, and P.~K. Maiti, ``Molecular mechanism of water
  permeation in a helium impermeable graphene and graphene oxide membrane,''
  {\em Physical Chemistry Chemical Physics}, vol.~17, no.~32, pp.~20557--20562,
  2015.

\end{thebibliography}
\end{document}